\documentclass[12pt,a4paper]{article}

\usepackage{graphicx}
\usepackage{epstopdf}
\usepackage{amsmath}
\usepackage{amsfonts}
\usepackage{amssymb}
\usepackage{amsthm}
\usepackage{slashed} 

\newcommand{\U}{{\mathrm U}}
\newcommand{\SO}{{\mathrm SO}}

\newcommand{\dd}{{\mathrm d}}      

\newcommand{\R}{{\mathbb R}}   
\newcommand{\C}{{\mathbb C}}   
\newcommand{\Z}{{\mathbb Z}}   

\numberwithin{equation}{section}

\DeclareMathOperator{\sgn}{sgn}

\DeclareMathOperator{\Realpart}{Re}

\newcommand{\bah}[1]{{\overline{#1}}}  

\theoremstyle{definition}

\newtheorem*{proposition}{Proposition}

\begin{document}

\title{A topological state sum model for fermions on the circle}

\author{John W. Barrett, Steven Kerr and Jorma Louko 
\\ \\
School of Mathematical Sciences\\
University of Nottingham\\
University Park\\
Nottingham NG7 2RD, UK\\
\\
}


\maketitle

\begin{abstract} A simple state sum model for fermions on a 1-manifold is constructed. The model is independent of the triangulation and gives exactly the same partition function as the Dirac functional integral with zeta-function regularisation. Some implications for more realistic physical models are discussed.
\end{abstract}


\section{Introduction}

Discrete models are used in physics as a practical way of defining a functional integral. In lattice gauge theories one obtains a model that approximates the continuum theory better as the number of discrete lattice cells is increased. In the continuum limit, the gauge theory should of course regain the full rotational and translational symmetry.

Topological quantum field theories can be defined using discrete models called state sum models~\cite{QGTQFT}. In this case the partition function is independent of the number of discrete cells (the simplexes of a triangulation). Similar models are used in quantum gravity, but the requirement that the partition function is independent of the triangulation is generally relaxed; however the independence must at least hold in a suitable limit in which the model approximates general relativity. 

Realistic quantum gravity models should of course include matter fields and so it is important to investigate models with matter fields that are independent of the triangulation, either exactly, or in a limit. This is our motivation for studying state sum models with fermionic fields. 

In this paper a very simple one-dimensional state sum model is presented in which a fermionic field is coupled to a background gauge field. A simple formula for the partition function of this model on a triangulated circle (i.e., a polygon) is presented in Section~\ref{ssmsection}. It is demonstrated that the partition function is independent of the triangulation and depends only on the holonomy of the gauge field. 

A heuristic argument is presented in Subsection \ref{wkfsection} that
shows that the state sum model has an action that is a discretisation
of the continuum Dirac action for a massless fermion field coupled to
the gauge connection. Then, in Subsection \ref{sectionfuncint} and the
two appendices, the partition function of this functional integral is
calculated precisely to show that it is in fact exactly the same as
the partition function of the state sum model.

The results are extended in Subsection \ref{masssection} to models with a mass term. A state sum model with a discretised mass term is presented. However in this case the model does depend on the number of edges $N$ in the triangulation of the circle. It is shown that as $N\to\infty$ it converges to a limit, so that the triangulation-independence is recovered in this limit. The limiting value is closely related to the Dirac functional integral with a mass term, differing only by a certain phase factor. Finally, the paper ends in Section \ref{discsection} with a discussion of the results, including a comparison with operator cut-off regularisations of the Dirac functional integral.

The results presented here complement previous work constructing
quantum gravity state sum models with fermion fields in dimension
three \cite{Fairbairn:2006dn,Dowdall:2010ej} and four
\cite{Bianchi:2010bn,Han:2011as}. These works construct discrete
analogues of the continuum Dirac functional integral according to the
heuristic continuum limit, as considered here in
Subsection~\ref{wkfsection}, but do not have a direct comparison with
the partition function of the continuum functional integral. The
results presented here give the first precise comparison of a discrete
fermionic model with the partition function of the continuum
functional integral.

It is an interesting question as to whether our results can be generalised to a higher-dimensional model. Some related properties of the Dirac operator with a gauge field on a graph have been studied previously~\cite{bolte}; this suggests there may also be an extension of our state sum model to graphs.

\section{The state sum model}
\label{ssmsection}

\subsection{Fermionic variables}

In a functional integral, the fermion fields anticommute and so are elements of a Grassmann algebra. Therefore the state sum model is based on a finite-dimensional Grassmann algebra over $\C$. Such an algebra is determined by a number of generators $a_1,a_2,\ldots,a_l$ satisfying the relations 
\begin{align}
a_i a_j+a_j a_i=0.
\end{align}

A function of the generators $f(a_1,\ldots, a_l)$ is a polynomial, which terminates at the highest monomial $a_la_{l-1}\ldots a_1$ due to the anticommutation. The Berezin integral
\begin{align}
\int\dd a_1\,\dd a_2\ldots\dd a_l\; f
\end{align}
is defined to be the coefficient of $a_la_{l-1}\ldots a_1$ in the expansion of $f$. No independent meaning is attached to the differentials in this formula, and they do not appear outside an integral. However the order of them in the integral is important; transposing two neighbouring differentials in the notation changes the sign of the integral.

It is possible to extend the above definitions to integration over a subset of coordinates, and perform the integral iteratively. So if $f=a_ka_{k-1}\ldots a_1 b$, where $b$ is a polynomial in the remaining variables $a_{k+1},\ldots, a_l$, then the integral is
\begin{align}
\int\dd a_1\,\dd a_2\ldots\dd a_k\; f= b,
\end{align}
with terms multiplying lower degree monomials in $a_1\ldots a_k$ integrating to zero. An example of iteration is the formula
\begin{align}
\int\dd a_1\,\dd a_2\; f=\int\dd a_1\,\left(\int\dd a_2\; f\right).
\end{align}

The integration enjoys translational invariance: if $c$ is any polynomial in the generators excluding $a_1$, then
\begin{align}
\int f(a_1+c)\,\dd a_1=\int f(a_1)\,\dd a_1.
\end{align}

In the applications of interest here the generators of the Grassmann algebra occur in pairs $a_i$, $b_i$  and are also grouped into a number of $n$-dimensional vectors, such as
\begin{align}
\psi=(a_1,a_2,\ldots a_n),\quad \bah\psi=(b_1,b_2,\ldots,b_n).
\end{align}
In this case the integral is defined with the notation
\begin{align}
\int \dd \psi\, \dd \bah\psi = \int \dd a_1\, \dd b_1\, \dd a_2\, \dd b_2 \ldots \dd a_n\, \dd b_n.
\end{align} 

Let $M$ be an $n\times n$ matrix with entries in $\C$. The basic gaussian integral is
\begin{align}
\int \dd\psi\, \dd\overline{\psi} \, e^{\overline{\psi} M \psi}   = \det M,  
\label{basicgaussian}  
\end{align}
which is proved by expanding the exponential. 

The result \eqref{basicgaussian} can be extended to the case of fermionic
source terms \cite{negele-orland}. Take $\overline{c}, d$ to be $n$-component vectors with
Grassmann-valued entries that are polynomial of odd degree in the
remaining generators (i.e., excluding components of both $\psi$ and
$\bah\psi$), and $M$ now an invertible matrix. 
We then have
\begin{align}  
\int \dd\psi\, \dd\overline{\psi} \, 
e^{\overline{\psi} M \psi + \overline{c} \psi + \overline{\psi} d} 
= \det M \, e^{-\overline{c} M^{-1} d}, 
\label{gaussian}
\end{align}
which is proved by first completing the square with the translations 
\begin{subequations}
\begin{align}
\overline{\psi}&\mapsto \overline{\psi} - \overline{c} M^{-1}, \\ 
\psi&\mapsto \psi -  M^{-1}d,
\end{align}
\end{subequations} 
and then using~\eqref{basicgaussian}.

\subsection{Definition of the state sum model}

The construction of the fermionic state sum model is based on fermionic variables at a discrete set of points. For this, a number of $n$-dimensional vectors $\psi_i$, $\bah\psi_i$ are used (with $i$ labelling the vectors, not the components). The Grassmann algebra is the one generated by all of the components of all of the vectors.

\begin{figure}[h!]  
\begin{center}
\includegraphics[scale=0.7]{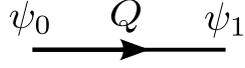}
\end{center} 
\caption{The partition function of an edge. The arrow indicates the orientation of the edge.}\label{fig1}
\end{figure}

The basic construction is the partition function for the interval $[0,1]\subset\R$ decorated with a fixed $n\times n$ matrix 
$Q$, as depicted in Figure~\ref{fig1}.
The partition function then is
\begin{align}
{{\mathbb Z}}^Q_{[0,1]}= e^{-\overline{\psi}_0 Q \psi_1}. \label{interval}
\end{align}
This has fermionic variables $\bah\psi_0$, associated to the boundary point $0$, and $\psi_1$ associated to~$1$.

Gluing two such partition functions together is carried out using the following proposition, which states that one can multiply matrices by the use of Berezin integration.
\proposition 
\begin{align}
\int \dd\psi_1 \dd\overline{\psi}_1 e^{-\overline{\psi}_0 Q_1 \psi_1} e^{\overline{\psi}_1 \psi_1} e^{-\overline{\psi}_1 Q_2 \psi_2} = e^{-\overline{\psi}_0 Q_1 Q_2 \psi_2}. \label{proposition1}
\end{align}

\begin{proof}
We have 
\begin{align}
\int \dd\psi_1 \dd\overline{\psi}_1\, e^{-\overline{\psi}_0 Q_1 \psi_1} e^{\overline{\psi}_1 \psi_1} e^{-\overline{\psi}_1 Q_2 \psi_2} &= \int \dd\psi_1 \dd\overline{\psi}_1 e^{\overline{\psi}_1 I \psi_1 - (\overline{\psi}_0 Q_1)\psi_1 - \overline{\psi}_1 (Q_2 \psi_2)},
\end{align}
where $I$ is the $n\times n$ identity matrix. 
The result \eqref{proposition1} follows using 
\eqref{gaussian} with
\begin{align}
\overline{c}=-\overline{\psi}_0 Q_1
, \ \ 
d=-Q_2 \psi_2
, \ \ 
M=I.
\end{align}
\end{proof}
Our interpretation of formula \eqref{proposition1} is that there is a bilinear form on the fermionic states,
\begin{align}
(f,g)=\int \dd\psi_1 \dd\overline{\psi}_1\, f(\psi_1)\, e^{\overline{\psi}_1 \psi_1} g(\bah{\psi}_1),  
\end{align}
and using this bilinear form to glue the partition functions results in
\begin{align}
\label{glueing}
{{\mathbb Z}}^{Q_1}_{[0,1]}{{\mathbb Z}}^{Q_2}_{[0,1]}={{\mathbb Z}}^Q_{[0,1]},
\end{align}
with $Q=Q_1Q_2$.

Clearly this can be iterated for the multiplication of 
any finite number of matrices, yielding
the definition of the state sum model. Explicitly, 
\begin{align}
\int \dd\psi_1\, \dd\overline{\psi}_1 
\ldots \dd\psi_{N-1}\, &\dd\overline{\psi}_{N-1} \; e^{-\overline{\psi}_0 Q_1 \psi_1} e^{\overline{\psi}_1 \psi_1} 
e^{-\overline{\psi}_1 Q_2 \psi_2} 
\ldots 
e^{-\overline{\psi}_{N-1} Q_N \psi_N} \notag \\
&= e^{-\overline{\psi}_0Q\psi_N}={{\mathbb Z}}^Q_{[0,1]} 
\label{repeated Berezin integral} 
\end{align}
with $Q=Q_1Q_2\ldots Q_N$.

The left-most expression in \eqref{repeated Berezin integral} 
is to be interpreted as the definition of the fermionic state sum model on an oriented interval that is triangulated using $N$ 1-simplexes. The equation itself is a statement that the partition function is independent of the triangulation, though of course provided the matrices $Q$ and the orientation match. Geometrically, if the $Q_i$ are all invertible, they may be viewed as the parallel transport operators for a connection 1-form on the interval. Then the partition function can be viewed as a function of the connection that is invariant under gauge transformations in the interior of the interval. 

The state sum model on the left-hand side of \eqref{repeated Berezin integral} can be modified to include observables, that is, non-trivial functions of the intermediate variables $\psi_i$, $\bah\psi_i$. In this sense the model is richer than the evaluation of the partition function on the right-hand side. The physical content of the model will be investigated further in Section \ref{fisection} by comparing it to a functional integral. 

\begin{figure}[h!]  
\begin{center}
\includegraphics[scale=0.7]{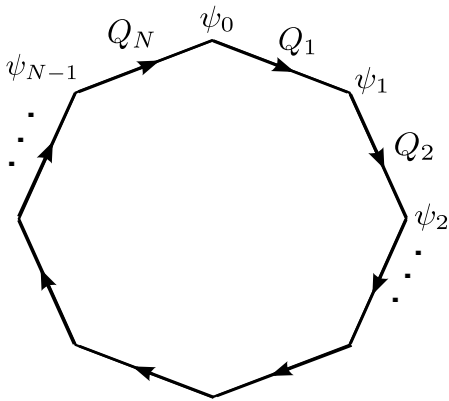}
\end{center} 
\caption{The state sum model on a triangulated circle.}\label{fig2}
\end{figure}

The partition function for the circle can be computed 
by gluing together the endpoints of the interval, as depicted in Figure~\ref{fig2}.

Mathematically, this is done by including an extra factor
$e^{\overline{\psi}_N \psi_N}$ in the integrand, identifying
$\psi_0=\psi_N$, $\overline{\psi}_0=\overline{\psi}_N$, and
integrating over the newly introduced variables,
\begin{align}
{{\mathbb Z}}_{\mathrm S^1}^Q
&=  \int \dd\psi_N \dd\overline{\psi}_N\,  
e^{\overline{\psi}_N \psi_N} e^{-\overline{\psi}_N  Q\psi_N} 
= \det \left(  I - Q\right), 
\label{circle} 
\end{align}
where the last equality follows from 
\eqref{basicgaussian} and the observation that $\overline{\psi}_N \psi_N$ and 
$\overline{\psi}_N  Q\psi_N$ commute. 

An immediate consequence of \eqref{circle} 
is that ${{\mathbb Z}}_{\mathrm S^1}^Q$ 
vanishes if $Q$ has an eigenvalue equal to~$1$, 
as is for example the case for the gauge group $\SO(2n+1)$. 
For $\SO(2n)$, ${{\mathbb Z}}_{\mathrm S^1}^Q$ is real-valued,
as the eigenvalues occur in complex conjugate pairs. 
There are however gauge groups for which 
${{\mathbb Z}}_{\mathrm S^1}^Q$ is complex-valued. 
An example of particular interest is~$\U(1)$, for which we have $Q = e^{-i\theta}$ with $\theta\in\R$, and so
\begin{align}
\U(1) : \quad 
{{\mathbb Z}}_{\mathrm S^1}^Q= 1 - e^{-i\theta}. 
\label{UoneZ}
\end{align}

This example shows that the partition function depends on the orientation of the
circle for some gauge groups. Given a circle with a connection, if the
holonomy around one direction is the matrix~$Q$, the holonomy around
the opposite direction is~$Q^{-1}$. For $\SO(n)$, $\det \left( I - Q\right) = \det
\left( I - Q^{-1}\right)$ for all~$Q$, and so the partition function is independent of orientation. For $\U(n)$, however $\det \left( I - Q\right)$ is the complex conjugate of 
$\det\left( I - Q^{-1}\right)$, and the $\U(1)$ example \eqref{UoneZ} shows that the partition function is sensitive to the orientation for generic $Q$.

\subsection{Gauge transformations}

Gauge transformations act on the partition function of $[0,1]$ by a linear transformation acting on each set of fermionic variables. Thus if $U_0$, $U_1$ are invertible matrices then the transformation is $\psi_1'= U_1\psi_1$ and $\overline{\psi}'_0= (U_0)^{-1T}\overline\psi_0$, using the inverse transpose matrix. The matrix $Q$ transforms as $Q' =  U_0 Q U_1^{-1}$. It is clear then that the partition function for the interval \eqref{interval} is invariant under gauge transformations,
\begin{equation}
  Z^{Q'}\bigl(\overline{\psi}'_0,\psi_1'\bigr)=Z^Q\bigl(\overline{\psi}_0,\psi_1\bigr).\end{equation}
 Similarly, the partition function of the circle \eqref{circle} is gauge-invariant.

Nothing so far has suggested that a variable $\psi_i$ is related to $\overline{\psi}_i$. However if, as the notation suggests, one regards these as complex conjugate variables, then the corresponding transformations should be complex conjugate matrices, $(U_i)^{-1T}=\overline U_i$, which means that these matrices are unitary.

\section{Comparison with the functional integral}
\label{fisection}

The partition function for the circle with a connection is compared to the corresponding functional integral in this section. First it is shown, in a heuristic fashion, that increasing the number of 1-simplexes can be interpreted as the state sum integrand converging to the exponential of the Dirac action for a fermion field. Then a precise evaluation of the Dirac functional integral is compared to the state sum model.
 
\subsection{Continuum limit of the action}
\label{wkfsection}

This subsection demonstrates that the state sum model is a discrete
version of the Dirac functional integral. This is done by looking at
the action for the state sum model and taking the limit $N\to\infty$,
assuming that the discrete values of the field sample values of a
differentiable field $\psi(t)$. The result of this heuristic limiting
process is the continuum Dirac action. 


The state sum model for the partition function on a circle triangulated with $N$ edges is
\begin{align}
 {{\mathbb Z}}_{\mathrm S^1}^Q = \int \left( \prod_{j=1}^{N} \dd\psi_j\, \dd\overline{\psi}_j \right) e^{-\overline{\psi}_N Q_1 \psi_1} e^{\overline{\psi}_1 \psi_1} ...  e^{\overline{\psi}_{N-1} \psi_{N-1}} e^{-\overline{\psi}_{N-1} Q_{N} \psi_{N}}e^{\overline{\psi}_N \psi_N}.
\label{circlessm} 
\end{align}
The integrand 
can be rewritten as
\begin{align}
J &=  \left( e^{\overline{\psi}_N \psi_N} e^{-\overline{\psi}_N Q_1 \psi_1} \right) \left( e^{\overline{\psi}_1 \psi_1} e^{-\overline{\psi}_1 Q_2 \psi_2} \right) 
\ldots  \left( e^{\overline{\psi}_{N-1} \psi_{N-1}} e^{-\overline{\psi}_{N-1} Q_N \psi_N} \right) 
\notag 
\\
&= e^{ i \Delta t \sum_{j=0}^{N-1} \overline{\psi}_j   i \left(  \frac{Q_{j+1} \psi_{j+1}-  \psi_j}{\Delta t}   \right) },
\end{align}
with $\psi_{N} =\psi_0$. 

With the choice $\Delta t = l/N$, where $l$ is a positive
constant, this looks precisely like a discretisation of the Dirac
action on a circle of circumference~$l$. Let $t$ be the coordinate on
the circle with range $0$ to~$l$. Then the $j$-th vertex corresponds
to the coordinate $t=jl/N$, and $\Delta t$ is the distance between
neighbouring vertices.

The limit $N\to\infty$ is the limit $\Delta t \rightarrow 0$. In this
limit the parallel transport of the connection is $Q_j = 1-i A^a(t)
X_a \Delta t+O(\Delta t)^2$, where $iX_a$ are the generators of the
relevant Lie algebra and $A^a(t)$ are the components of the connection
1-form.

Assuming the field values are differentiable, we have 
\begin{align}
\label{derivative}
\lim_{\Delta t \rightarrow 0} i \left(  \frac{Q_{i+1} \psi_{i+1}-  \psi_i}{\Delta t}   \right) 
&=  \slashed{D} \psi(t), 
\end{align}
where the gauge covariant Dirac operator $\slashed{D}$ is given by 
\begin{align}
\slashed{D}=i \frac{d}{dt} + A^a(t) X_a . 
\label{diracoper-gen}
\end{align}
The single gamma matrix is equal to the complex number~$i$. 
There is no spin connection contribution to \eqref{diracoper-gen}
because the tangent space is one-dimensional and the Lie algebra $\mathfrak{so}(1)$ is trivial.  

In the limit $\Delta t \rightarrow 0$ the sum converges to an integral, 
\begin{align} 
\label{integral}
\lim_{\Delta t \rightarrow 0} \Delta t \sum_{j=0}^{N-1}f(j\Delta t) 
= \int_0^l f\,\dd t ,
\end{align}
giving the continuum limit of the state-sum integrand as
\begin{align}
\lim_{\Delta t \rightarrow 0} J 
\, = \, 
e^{ i \int_0^l \dd t\; \overline{\psi}(t)  \slashed{D} \psi(t)}.
\end{align}

\subsection{The functional integral}
\label{sectionfuncint}

In this subsection the above heuristic considerations are confirmed by
an explicit and precise evaluation of the functional integral in the $\U(n)$ case.

The continuum partition function is  
\begin{align} 
{\mathbb F}^A_{S^1} = \int \mathcal{D} \psi \mathcal{D} \overline{\psi}  \, 
e^{i\int_0^{l} \dd t \, \overline{\psi}(t) \slashed{D} \psi(t)} , 
\label{Dirac generating functional} 
\end{align}
where $\slashed{D}$ is given by \eqref{diracoper-gen} 
and the fermions obey periodic boundary conditions. These boundary conditions can be regarded as the choice of the `Lie group' spin structure for the circle \cite{Kirby}. 
The functional integral in \eqref{Dirac generating functional} is defined 
via zeta-function
regularisation~\cite{baer-schopka,vassilevich-manual}, as reviewed in
Appendix~\ref{app:continuumdetreg}. The result is 
\begin{align}
{\mathbb F}^A_{S^1} = \det (i\slashed{D}) 
= e^{ i \frac{\pi}{2} \eta_{\slashed{D}} (0) } e^{ - \frac{1}{2} \zeta_{\slashed{D}^2}' (0)} , 
\label{determinant definition} 
\end{align}
where the functions $\eta_{\slashed{D}}$ and $\zeta_{\slashed{D}^2}$ 
are defined in Appendix~\ref{app:continuumdetreg}. 

It is shown in Appendix \ref{appsec:evaluation} 
that for $\U(1)$  
\begin{align}
{\mathbb F}^A_{S^1}={{\mathbb Z}}_{\mathrm S^1}^Q,
\label{surprising}
\end{align} 
where $Q$ is the holonomy of the connection~$A$. The result
\eqref{surprising} generalises immediately to $\U(n)$ by diagonalising
the connection with a gauge transformation, whereupon the functional
integral is the product of a number of $\U(1)$ functional
integrals. The Dirac functional integral is invariant under these
gauge transformations because it depends only on the eigenvalues of
the Dirac operator, which are gauge invariant.

The result \eqref{surprising} is surprising because the eigenvalues of
the Dirac operator are unbounded and so the na\"\i{}ve determinant of the
Dirac operator, the product of its eigenvalues, diverges. Somehow the
discrete model both approximates the eigenvalues of the continuum
operator yet also avoids the divergence, and miraculously imitates the
zeta-function regularisation.

Some insight into the result \eqref{surprising} can be gained by
comparing the eigenvalues for the continuum Dirac operator with the
eigenvalues for its discrete version.

The discrete version of the Dirac operator is a matrix $M$ acting on
the vectors $\psi = \bigoplus_{j=1}^{N} \psi_j$. It is determined by
\begin{align}
\overline{\psi} iM \psi= \sum_{j=0}^{N-1} \overline{\psi}_j  
\left(  \psi_j -  Q_{j+1} \psi_{j+1} \right) ,
\end{align}
and can be written in block form as 
\begin{align} 
iM =
\left( \begin{array}{ccccc}
 1 & -Q_2 & & & \\
 & 1 & -Q_3 & &  \\
 & & \ddots & \ddots & \\
 & & & 1 & -Q_N \\
 -Q_1 & & & & 1 \end{array} \right).
\end{align}
where each $Q_i$ is in $\U(n)$.  Since we have seen that the partition
function associated with $M$ is exactly the same as the continuum partition function,
it must be that the matrix $M$ is in some sense approximating the
differential operator~$\slashed{D}$. We now make this more explicit.

For simplicity, we specialise to~$\U(1)$. The 
eigenvalues of $iM$ are $\mu_k =
1-\alpha_k$, where $\alpha_k$ are the $N$ roots of  
\begin{align}
Q=\alpha^N  , 
\end{align}
and the corresponding eigenvectors are 
\begin{align} 
\left( \begin{array}{ccccc}
Q^{-1}_1 \\
\alpha_k Q^{-1}_2 Q^{-1}_1  \\
\alpha_k^2 Q^{-1}_3 Q^{-1}_2 Q^{-1}_1\\
\ldots \\
\alpha_k^{N-1} Q^{-1}_{N}\ldots Q^{-1}_1 
\end{array} \right).
\end{align}
Taking $Q=e^{-i\theta}$ with $\theta \in [0,2\pi)$, as in~\eqref{UoneZ}, 
we have 
\begin{align}
\mu_k = 1-e^{-i\left( \frac{\theta +2k\pi}{N} \right)},
\label{eq:disc-evalues}
\end{align}
where the distinct eigenvalues are obtained 
by selecting a suitable set of distinct values of~$k$. For example, this can be done explicitly by taking  
$k=[(1-N)/2],\ldots,[(N-1)/2]$, where $[x]$ stands for the largest integer 
that is less than or equal to~$x$. 

To compare with the continuum Dirac operator, 
we note that the eigenvalues \eqref{eq:disc-evalues} 
of $iM$ have the large $N$ expansion 
\begin{align}
iM: \quad \mu_k &= 1-e^{-i\left( \frac{\theta +2\pi k}{N} \right)} 
 =i\left(\frac{\theta +2\pi k}{N}\right)
+O\left(\left(\frac{\theta +2\pi k}{N}\right)^2\right) , 
\label{eq:disc-evalues-expansion}
\end{align}
while in Appendix \ref{appsec:evaluation}
it is shown that the eigenvalues of 
$i\slashed{D}$ are
\begin{align}
i\slashed{D}: \quad \mu_k 
= i\left(\frac{\theta+2\pi k}{l}\right) , 
\quad k \in\Z. 
\label{eq:cont-evalues}
\end{align}
The expressions \eqref{eq:disc-evalues-expansion} and \eqref{eq:cont-evalues}
coincide modulo $O(N^{-2})$ if the circle has length $l=N$ and $k$ is held
fixed. The matrix $iM$ hence approximates the operator $i\slashed D$ in
the sense that the eigenvalues of small modulus 
coincide in the limit of a large circle. 

Note from \eqref{eq:disc-evalues-expansion} that the eigenvalues of
$M$ are complex but the imaginary part is subdominant as $N\to\infty$ with
fixed~$k$.  This raises the question as to whether it is fruitful to
think of $M$ as a cut-off version of~$\slashed D$. We shall return to
this question in Section~\ref{discsection}.

\subsection{Mass term}
\label{masssection}

In this subsection we show that inclusion of a mass term breaks triangulation independence, but gives a well-defined partition function as $N\to\infty$.
We start now with a massive continuum fermionic action
\begin{align}
S = \int_0^l \dd t\; \overline{\psi}(t) \left(  \slashed{D} - m \right) \psi(t),
\end{align}
and we follow the usual discretisation procedure using \eqref{derivative} and~\eqref{integral}. The state sum is given by
\begin{align}
{\mathbb I}_{\mathrm S^1} =  \int \left( \prod_{j=0}^{N-1} d\psi_j d\overline{\psi}_j \right) e^{ \sum_{j=0}^{N-1} \overline{\psi}_j  \left(  (1-im\Delta t) \psi_j -  Q_{j+1} \psi_{i+1} \right) },
\end{align}
where $\Delta t = l/N$, $\psi_{N} = \psi_0$, and the mass term has been discretised according to 
\begin{align}
\int_0^l m\overline{\psi} \psi\;\dd t \rightarrow m\Delta t \sum_{j=0}^{N-1}  \overline{\psi}_j \psi_j.  
\label{mass}
\end{align}

We can evaluate ${\mathbb I}_{\mathrm S^1}$ in a similar way as before to obtain
\begin{align}
{\mathbb I}_{\mathrm S^1} = \det \left((1-im\Delta t)^N - \prod_{i=1}^N Q_i\right),
\end{align}
which is clearly not triangulation independent. The triangulation dependence appears to have crept in with the introduction of a fixed mass scale~$m$. A way to regain triangulation independence might be to relegate $m$ from a fixed parameter to a function of $N$ and a new fixed positive parameter $m'$ by $1-im\Delta t = e^{-im' l /N}$; 
however, this would mean allowing $m$ to be complex-valued. 

With $m$ fixed, the limit $N \to \infty$ does however yield a well-defined answer,  
\begin{align}
\lim_{N \to \infty}  \det \left( (1-im\Delta t)^N - \prod_{i=1}^N Q_i \right)  = \det (e^{- i m l} - Q), 
\label{limit}
\end{align}
where $Q$ is the holonomy around the circle. This can be compared to the evaluation of the partition function with a mass term in the continuum $\U(n)$ case. The calculation is done identically but using $A' = A-m$. The answer comes out as
\begin{align}
{\mathbb F}^{A'}_{S^1} = \det( 1-e^{ 2\pi i m}Q), 
\end{align}
which differs from \eqref{limit} by a phase factor.

The disparity is due to the different way in which the gauge field and mass term are treated. In the continuum calculation they are bundled together into one term. However, in the discrete calculation, the gauge field is realised as parallel transporters while the mass term couples fields at the same point.

\section{Discussion}
\label{discsection}

In this paper, a fermionic state sum model on one-dimensional
manifolds with a connection is defined. The partition function of the
state sum model on the circle is the same as the functional integral
of a complex fermion field with the Dirac action and the usual
coupling to the connection gauge field.

A curious feature of the partition function for the circle is that it
does not depend on the length of the circle, whereas the eigenvalues
of both the continuum Dirac operator $\slashed{D}$ and the discrete
Dirac operator $M$ clearly do.

Some more insight into this can be gained by examining an operator cut-off regularisation of the functional integral. Such a regularisation depends on a cut-off scale $c$. It replaces $iD$ with $f(iD)$, where the function $f(z)$ of a complex variable is the identity function $f(z)=z$ for $|z|\ll c$, but $f(z)=1$ for $|z|\gg c$. Thus it effectively removes the eigenvalues of $D$ with magnitude above the cut-off $c$. An example of such a cut-off regularisation is the Schwinger proper time regularisation. 

As $c\to\infty$, the regularised determinant diverges. A calculation for the Schwinger proper time regularisation shows that the leading asymptotic term for $\log\det |\slashed{D}|$ is $(l/\pi)\,c\log c$. This can be confirmed in a simple way for special cases, e.g.\ $A=1/2$, using a sharp cut-off and Stirling's formula for the asymptotic expansion of the factorial.

The cut-off regularisation only agrees with the zeta function method once the leading asymptotic terms are removed \cite{baer-schopka}. It is worth noting that these leading divergent terms are proportional to $l$, thus explaining the role played by the length of the circle in evaluating this determinant. Thus to get a partition function that converges as $c\to\infty$, one can multiply the cut-off regularised determinant with a `cosmological term' 
$e^{\Lambda(c) l}$, choosing a suitable function $\Lambda(c)$ to renormalise the 
coefficient of $l$ as $c\to\infty$. This is the same as adding a term $\Lambda(c) l$ to the exponent in \eqref{Dirac generating functional} when using an operator cut-off regularisation. The functional integral formula for this regularisation is then  
\begin{align}
\int \mathcal{D} \psi \mathcal{D} \overline{\psi}  \, e^{i\int_0^{l} \dd t \, (\overline{\psi}(t) \slashed{D} \psi(t)-i\Lambda)}.
\end{align} 
Similar divergences do not occur with cut-off regularisations of the eta invariant in the phase term \cite{bismut-freed}, so there are no additional parameters to renormalise for the phase of the partition function.

Our conclusion from this discussion is that the state sum model is a more subtle regularisation of the determinant of the Dirac operator than a mere operator cut-off. 

Another aspect of this is the absence of the phenomenon of fermion doubling which commonly occurs with lattice discretisations of the Dirac action \cite{fermion doubling}. One of the standard assumptions of discussions of fermion doubling is that the action of the lattice theory is real (i.e. equal to its conjugate with a suitable definition of complex conjugation for fermionic variables). However the action used in the state sum model \eqref{repeated Berezin integral} is in fact not real, and the resulting complex nature of the eigenvalues of $M$ appears to be a crucial aspect of the model.

Finally, it is worth making a brief mention of possible generalisations. One can include an integration over the $\U(n)$ gauge field in the state sum model using the normalised Haar measure on the group 
\begin{align}
{{\mathbb T}}_{[0,1]}=\int\dd Q\; {{\mathbb Z}}^Q_{[0,1]} . 
\end{align}
The resulting partition function $\mathbb T_{[0,1]}$ is independent of any background structure;
\eqref{glueing} yields
\begin{align} {{\mathbb T}}_{[0,1]}{{\mathbb T}}_{[0,1]}={{\mathbb T}}_{[0,1]}\end{align}
so that ${{\mathbb T}}_{[0,1]}$ is a projection operator. The corresponding partition function for a circle is thus an integer, and can be interpreted in terms of counting states in the space of polynomials in the fermion variables at a point that are invariant under the group action.

Some aspects of the model presented here have been studied for the Dirac operator on a graph; the formula \eqref{limit} is studied in \cite{Carlson}. The most interesting generalisation would be to models in higher dimensions. Whilst one cannot expect such explicit evaluations as presented here for the circle, we hope that the one-dimensional model clarifies some of the issues that may arise in higher dimensions.

\section*{Acknowledgments}

We thank Sara Tavares for discussions. JWB and JL were supported in part by STFC. JWB and SK thank the QG research network programme of the European Science Foundation for support for the 
workshop on representation theoretical and categorical structures in quantum geometry and conformal field theory, Erlangen, October 2011, where this work was reported.

\begin{appendix}

\section{Appendix: Zeta-function regularisation of the 
Dirac determinant}
\label{app:continuumdetreg}

In this appendix, the zeta-function regularisation of $\det
D$ for a self-adjoint operator $D$ that is not
necessarily positive definite \cite{baer-schopka} is reviewed. This regularisation is then adapted to~$\det (iD)$, which is compared with $\det D$.

\subsection{Positive definite $D$}

To begin, suppose $D$ is a Hermitian, strictly positive operator in a
finite-dimensional Hilbert space.  The zeta-function $\zeta_{D} (s)$
of $D$ is defined for $s\in\C$ by
\begin{align}
\zeta_{D} (s) = \sum_k \frac{1}{\lambda_k^s},
\label{eq:zetaLdef}
\end{align}
where $\lambda_k$ are the eigenvalues of~$D$. 
As the eigenvalues are finitely many and positive,  
$\zeta_{D} (s)$ is well-defined and holomorphic in~$s$. 
An elementary computation shows that 
\begin{align}
\det D = \prod_k \lambda_k = e^{- \zeta_D'(0) } . 
\label{eq:detLdef}
\end{align}

The point of \eqref{eq:detLdef} is that the rightmost expression can
be adopted as the definition of $\det D$ even when the Hilbert space
is infinite dimensional, provided the spectrum of $D$ is still
discrete and the sum in \eqref{eq:zetaLdef} converges for sufficiently
large $\Realpart s$ to define $\zeta_{D} (s)$ as a function that can
be analytically continued to $s=0$.  The analytic continuation in $s$
provides a prescription that removes the infinity from the formally
divergent product~$\prod_k \lambda_k$.

\subsection{Indefinite $D$}

Suppose that $D$ is an indefinite Hermitian operator in a
finite-dimensional Hilbert space, such that the spectrum of $D$ does
not contain zero.  We wish to express $\det D$ and $\det(iD)$ in a
form similar to~\eqref{eq:detLdef}.  The new issue is to accommodate
the negative and imaginary eigenvalues. 

Let $\lambda_k$ denote the eigenvalues of~$D$, enumerated so that
$\lambda_k>0$ for $k>0$ and $\lambda_k<0$ for $k\le0$.  
We define two zeta-functions by 
\begin{subequations}
\label{eq:threezetas-def}
\begin{align}
\zeta_{D,\epsilon} (s) 
&= \sum_{k>0} \frac{1}{\lambda_k^s}
+ e^{i\epsilon\pi s}\sum_{k\le0} \frac{1}{{(-\lambda_k)}^s} , 
\label{eq:zetasDeps-def}
\\
\zeta_{iD} (s) 
&= e^{-i\pi s/2}\sum_{k>0} \frac{1}{\lambda_k^s}
+ e^{i\pi s/2}\sum_{k\le0} \frac{1}{{(-\lambda_k)}^s}, 
\label{eq:zetasiDeps-def}
\end{align}
\end{subequations}
where $\epsilon \in \{1,-1\}$. 
For integer argument these functions agree with na\"\i{}vely 
allowing negative or imaginary eigenvalues in~\eqref{eq:zetaLdef}. 

We next define the eta-function of $D$ by 
\begin{align}
\eta_{D} (s) 
& = \sum_k \frac{ \sgn \lambda_k}{ {| \lambda_k |}^s}
= \sum_{k>0} \frac{1}{\lambda_k^s} - \sum_{k\le0}
\frac{1}{{(-\lambda_k)}^s} . 
\label{eq:etainv-def}
\end{align}
Finally, since 
$D^2$ is positive definite, its zeta-function is
defined by the replacements 
$D\to D^2$ and $\lambda_k \to \lambda_k^2$
in~\eqref{eq:zetaLdef}. It follows that 
\begin{align}
\zeta_{D^2} (s/2) 
&= \sum_k \frac{1}{{|\lambda_k|}^s}
= \sum_{k>0} \frac{1}{\lambda_k^s} + \sum_{k\le0} \frac{1}{{(-\lambda_k)}^s} . 
\label{eq:zetasD2}
\end{align}

All the functions in \eqref{eq:threezetas-def}, \eqref{eq:etainv-def}
and \eqref{eq:zetasD2} are well-defined and holomorphic. They satisfy
\begin{subequations}
\label{eq:threezetas-2nd}
\begin{align}
\zeta_{D,\epsilon}(s) 
& = \frac12 \! \left(1 + e^{i\epsilon\pi s} \right) \zeta_{D^2} (s/2) 
+ \frac12 \! \left(1 - e^{i\epsilon\pi s} \right) \eta_{D} (s) , 
\label{eq:zetasDeps-2nd}
\\[1ex]
\zeta_{iD} (s) 
& = \cos(\pi s/2) \zeta_{D^2} (s/2) 
- i \sin(\pi s/2) \eta_{D} (s) , 
\label{eq:zetasD-2nd}
\end{align}
\end{subequations}
and differentiation of \eqref{eq:threezetas-2nd} at $s=0$ yields 
\begin{subequations}
\label{eq:threezetas-diff}
\begin{align}
\zeta'_{D,\epsilon} (0) 
& = \frac12 \zeta'_{D^2} (0) 
+ \frac{i\epsilon\pi}{2} 
\bigl( 
\zeta_{D^2} (0) -  \eta_{D} (0) \bigr) , 
\\[1ex]
\zeta'_{iD} (0) 
& = \frac12 \zeta'_{D^2} (0) 
- \frac{i\pi}{2} \eta_{D} (0) . 
\end{align}
\end{subequations}

We are now ready to turn to the determinants. 
An elementary computation using
\eqref{eq:threezetas-def} shows that 
\begin{subequations}
\begin{align}
\det D &= \prod_k \lambda_k = e^{- \zeta_{D,\epsilon}'(0) } , 
\label{eq:detsDepsdef}
\\
\det(iD) &= \prod_k (i\lambda_k) = e^{- \zeta_{iD}'(0) } . 
\label{eq:detisDdef}
\end{align}
\end{subequations}
Using \eqref{eq:threezetas-diff}, we thus have 
\begin{subequations}
\label{eq:detsbothfinal}
\begin{align}
\det D &= 
e^{ i\epsilon \frac{\pi}{2} \left(\eta_{D} (0) - \zeta_{D^2} (0) \right)} 
\, e^{ - \frac{1}{2} \zeta_{D^2}' (0)} , 
\label{eq:detsDepsfinal}
\\
\det(iD) &= 
e^{ i \frac{\pi}{2} \eta_{D} (0) } \, e^{ - \frac{1}{2} \zeta_{D^2}' (0)} . 
\label{eq:detisDfinal}
\end{align}
\end{subequations}

Formulas \eqref{eq:detsbothfinal} provide definitions for
$\det D$ and $\det(iD)$ that extend to the case when the Hilbert space is
infinite-dimensional and separable and the spectrum of $D$ is discrete
with suitable asymptotic properties. The values $\eta_{D} (s)$ and
$\zeta_{D^2} (s/2)$ are in this situation 
defined by \eqref{eq:etainv-def}
and \eqref{eq:zetasD2} for sufficiently large $\Realpart s$ and the functions are
analytically continued to $s=0$.

An important difference between $\det D$ and $\det(iD)$ arises from
the phases in~\eqref{eq:threezetas-def}.  In the definition
of $\zeta_{iD}$~\eqref{eq:zetasiDeps-def}, 
we chose the phases of the positive and
negative eigenvalue terms to be the opposite for real argument, with
the consequence that in the finite-dimensional case $\zeta_{iD}$ is
real-valued for real argument whenever the spectrum of $D$ is
invariant under $D \to -D$. In the definition of
$\zeta_{D,\epsilon}$~\eqref{eq:zetasDeps-def}, by contrast, 
the branch of ${(-1)}^{-s}$ in the negative eigenvalue terms
cannot be fixed by a similar symmetry argument, and we parametrised this
ambiguity by $\epsilon \in \{1,-1\}$, finding that 
$\epsilon$ still survives in the final formula 
\eqref{eq:detsDepsfinal} for~$\det D$. 
In the finite-dimensional case $\eta_{D} (0) - \zeta_{D^2} (0)$ is
an even integer, and the right-hand side of \eqref{eq:detsDepsfinal}
is hence independent of~$\epsilon$. In the infinite-dimensional case,
however, the two values of $\epsilon$ can yield different regularised
values for~$\det D$: an example will occur in 
Appendix~\ref{appsec:evaluation}. The choice $\epsilon=1$ is related
to our regularisation of $\det(iD)$ since $\zeta_{D,1}(s) = e^{i\pi
  s/2} \zeta_{iD}(s)$, whereas the formulas in \cite{baer-schopka}
make the choice $\epsilon=-1$.


\section{Appendix: Dirac determinant on the circle for~$\U(1)$}
\label{appsec:evaluation}

In this appendix we evaluate $\det(\slashed{D})$ and
$\det(i\slashed{D})$ for the Dirac operator $\slashed{D}$
\eqref{diracoper-gen} with the gauge group~$\U(1)$, using the
regularisation~\eqref{eq:detsbothfinal}.

For $\U(1)$, the Dirac operator \eqref{diracoper-gen} reduces to
$\slashed{D} = i\frac{d}{dt} + A(t)$, where $A$ is a real-valued
function of the coordinate $t\in[0,l]$, the Hilbert space is
$L^2\bigl([0,l]\bigr)$, and both $A$ and the domain of $\slashed{D}$
have periodic boundary conditions.  By a gauge transformation we may
assume $A$ to take a constant value, which we denote by $2\pi a/l$,
and by a further gauge transformation we may assume $a\in[0,1)$.  The
holonomy of $A$ is $Q = e^{-2\pi i a}$. Note that $a$ is uniquely
determined by the holonomy.

Solving the eigenvalue equation $\slashed{D}\psi = \lambda\psi$,
subject to the $l$-periodicity, shows that the eigenvalues are
$\lambda_k = 2\pi(k+a)/l$ where $k \in \Z$. We exclude the special
case $a=0$, in which one eigenvalue vanishes.  We then have
$a\in(0,1)$, all the eigenvalues are non-vanishing, and we are in the
situation covered by Appendix~\ref{app:continuumdetreg}. 

Consider first $\zeta_{\slashed{D}^2}$. Assuming $\Realpart s > 1$, we have 
\begin{align}
\zeta_{\slashed{D}^2} (s/2) &= 
{(2\pi/l)}^{-s}
\sum_{k \in \Z} \frac{1}{{|k+a|}^{s}} 
\notag 
\\
&= 
{(2\pi/l)}^{-s}
\left(\sum_{j=0}^{\infty} \frac{1}{{(j+a)}^{s}} 
+ \sum_{j=0}^{\infty} \frac{1}{{(j+1-a)}^{s}} \right)
\notag 
\\
&= 
{(2\pi/l)}^{-s}
\bigl( \zeta_H (s, a) + \zeta_H (s, 1-a) \bigr), 
\end{align}
where the sums are absolutely convergent and the 
Hurwitz zeta function $\zeta_H$ is defined by \cite{dlmf}
\begin{align}
\zeta_H (s,q) = \sum_{j=0}^{\infty} \frac{1}{{(j+q)}^s}.
\end{align}
Analytically continuing to $s=0$ and using 
25.11.13, \; 25.11.18 
and 5.5.3 in~\cite{dlmf}, we find 
\begin{subequations}
\label{eq:app2zzetas}
\begin{align}
& \zeta_{\slashed{D}^2} (0)=0 , 
\label{eq:app2zetanought}
\\
& e^{ - \frac{1}{2} \zeta_{\slashed{D}^2}' (0)} 
= 2 \sin \pi a .
\label{eq:appexpD2}
\end{align}
\end{subequations}

Consider now $\eta_{\slashed{D}}$. Assuming again $\Realpart s > 1$, we have 
\begin{align} 
\eta_{\slashed{D}} (s) &= 
{(2\pi/l)}^{-s}
\sum_k \frac{ \sgn (k+a)}{ {| k+a |}^s} 
\notag 
\\
&= 
{(2\pi/l)}^{-s}
\bigl(
\zeta_H (s,a) - \zeta_H(s,1-a) \bigr).
\end{align}
Analytically continuing to $s=0$ and using 25.11.13 in~\cite{dlmf}, 
we find 
\begin{align}
\eta_{\slashed{D}} (0) 
= 1-2a , 
\label{eq:appetainvfinal}
\end{align}
in agreement with \cite[\S 1.13]{gilkey}.

Using \eqref{eq:detsbothfinal} with \eqref{eq:app2zzetas}
and~\eqref{eq:appetainvfinal}, and recalling
$Q=e^{-2\pi i a}$, we have
\begin{subequations}
\begin{align}
\det \slashed{D}
&= 
\begin{cases}
1 - Q& \text{for $\epsilon=1$},\\
1 - Q^{-1}& \text{for $\epsilon=-1$}, 
\end{cases}
\label{eq:app:final-det-eps}
\\[1ex]
\det(i\slashed{D})
& = 
1 - Q . 
\label{eq:app:final-det}
\end{align}
\end{subequations}

Note that $\det \slashed{D}$ and $\det(i\slashed{D})$ depend only on
the holonomy and not on~$l$. The reason for the $l$-independence can
be traced to the property~\eqref{eq:app2zetanought}.

The modulus of the final result \eqref{eq:app:final-det} for
$\det(i\slashed{D})$ agrees with the calculation in the physics
literature of the ratio of two such determinants with different values
of $a$~\cite{paper with mistake}.  However the phase does not agree,
presumably due to the fact that the definition of this ratio in
\cite{paper with mistake} is given as an infinite product that is not
absolutely convergent.

Finally, it is worth noting that when $a=0$, the formulas \eqref{eq:app:final-det-eps} and \eqref{eq:app:final-det} all extend by continuity to give determinant equal to $0$. This is consistent with the fact that $\slashed{D}$  has $0$ as an eigenvalue.
\end{appendix}

\end{document}